\documentclass[aps,twocolumn,prb,superscriptaddress, showkeys]{revtex4}
\usepackage{amsfonts}
\usepackage[final,hiresbb]{graphicx}
\usepackage{amsmath}
\usepackage{amssymb}
\usepackage{amstext}

\usepackage{color}
\usepackage{braket}

\usepackage{mathrsfs}
\usepackage[cal=boondox]{mathalfa}

\definecolor{Green}{RGB}{0,204,102}
\definecolor{Magenta}{RGB}{220,0,220}
\definecolor{Blue}{RGB}{0,0,255}
\definecolor{Red}{RGB}{255,0,0}

\begin{document}

\title{Exciton Interferometry}  
\author{Ariel Shlosberg}
\affiliation{Department of Physics, Colorado School of Mines, Golden, CO 80401, USA}
\author{Mark T. Lusk}
\email{mlusk@mines.edu}
\affiliation{Department of Physics, Colorado School of Mines, Golden, CO 80401, USA}

\date{\today}

\begin{abstract}
An exciton beam splitter is designed and computationally implemented, offering the prospect of excitonic interferometry. Exciton interaction between propagation conduits is modeled using a coupling parameter that varies with position. In practice, this variation can be realized by a change in the distance separating conduits as would occur if they crossed at oblique angles. Two such excitonic beam splitters can be combined to comprise an excitonic analog to a Mach-Zehnder interferometer, allowing the relative phase shift between two signals to be used to tailor the output populations on each channel. In contrast to optical splitters, an excitonic signal can be coherently split among more than two channels. These ideas are computationally demonstrated within an idealized setting in which each site is idealized as a two-level system. Physical implementations include molecular and coupled cavity settings as well as combinations of these. This adds to the developing inventory of excitonic analogs to optical elements. 

\end{abstract}
\maketitle
\section{Introduction}

Photon absorption by semiconductors can generate an excitation in which an electron-hole exciton will subsequently propagate~\cite{PhysRevLett.89.097401}. Since the footprint and speed of such packets can be modified by structuring the medium as well as through external control, excitonic circuits can be designed to manipulate both packet information and energy content~\cite{PhysRevB.95.195423}. Excitons can interact with each other and their constituents can dissociate, adding an additional degree of richness to the dynamics\cite{PhysRevB.95.195423,Dostál2018}. Since it is possible to convert back and forth between electromagnetic and excitonic manifestations, excitonic logic is a promising facet of opto-electronic systems. This motivates an ongoing effort to develop excitonic analogs to key optical processing components. It is also a rich new arena for fundamental studies of quantum correlation. 

Perhaps one of the most useful pieces of optical hardware is the  beam splitter, which can be manifested classically as a semi-reflecting, lossless mirror~\cite{Degiorgio_1980}. From a circuit perspective, this amounts to a four-port element with output channels that are a linear combination of the two inputs. This enables interferometric measurements between two channels, a mainstay of optical sensing technologies. Optoexcitonic circuits can be used to implement solid state qubits and quantum logic gates \cite{Kim2013, PhysRevA.93.062342}. Moreover, coupled excitonic sites have broad application for quantum communication as they can function as Heisenberg spin chain systems \cite{PhysRevA.69.052315, PhysRevA.72.062326, doi:10.1080/00107510701342313, PhysRevLett.91.207901}. Beam splitters are also essential to many quantum information processing protocols such as quantum teleportation schemes, dense quantum coding, and entanglement distillation \cite{PhysRevA.65.052313, Bouwmeester1997, PhysRevLett.76.4656, Dong2008}. At a more fundamental level, beam splitters lie at the heart of studies at the foundations of quantum theory--e.g. ``which-path" investigations, EPR paradox violations, and the complementarity principle\cite{PhysRevLett.84.1, PhysRevLett.92.210403, PhysRevA.85.032121}. The design of an excitonic beam splitter, coupled with already existing excitonic component analogs, allows for an investigation of many of these protocols and experiments in a non-optical setting~\cite{excbell, PhysRevB.96.155104}. 

Due to the bosonic nature of excitons, generating Bose-Einstein condensates within solid-state systems and within indirect exciton gases is of interest. Exciton coherence has been investigated via photoexcitation and subsequent interferometric measurements on the light ultimately emitted\cite{Kasprzak2006, doi:10.1021/nl300983n}. However, interferometric confirmation of the spontaneous coherence of excitons has been shown to be reproducible as an effect of imaging partially coherent light while only taking into account the spatial distributions characteristic of the samples\cite{doi:10.1088/1361-6455}. This demonstrates a flaw in using photonic interferometry to categorize excitonic coherence in the case of condensate systems. 

\begin{figure}[hptb]
\begin{center}
\includegraphics[width=0.48\textwidth]{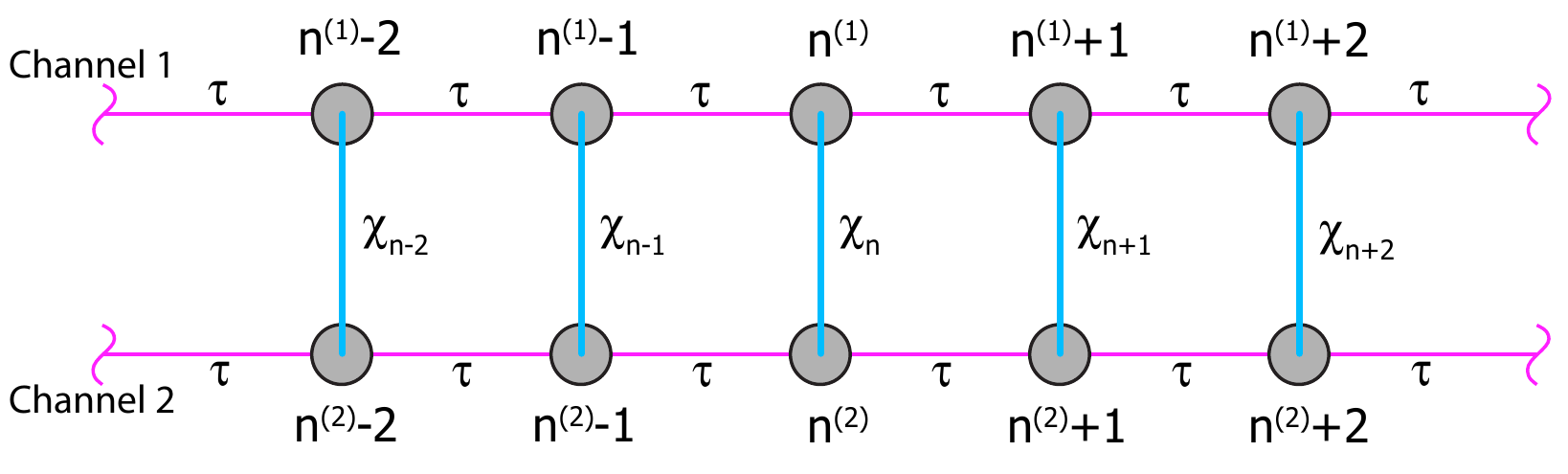}
\end{center}
\caption{\emph{Beam splitter schematic.} Two-channel excitonic beam splitter.} 
\label{Splitter_Schematic}
\end{figure}

It is therefore desirable to construct an analog to the optical beam splitter that can allow for direct interferometric sensing of excitonic wave-packets. A simple means of accomplishing this is to arrange two excitonic conduits so that they experience a degree of coupling over a limited spatial extent. This is illustrated in Figure \ref{Splitter_Schematic}, where each channel has a constant inter-site coupling, $\tau$, but there is also site-dependent inter-channel coupling, $\chi_n$. The individual channels can be idealized as being composed of discrete sites ($n^{(j)},\ j=1,2$), where excitons are localized to a site and the electron-hole pairs remain bound as quasi-particles---i.e. Frenkel excitons \cite{PhysRevB.92.241112}. Here a variation in the inter-channel coupling corresponds to an alteration in the spatial separation between the two channels or to the case where the two channels cross at an oblique angle. This inductive design for an excitonic beam splitter is analogous to experimental realizations of waveguide photonic beam splitters that rely on optical tunneling \cite{1253058, J_ns_2015}.

%

The beam splitting character of this excitonic circuit element can be explored using a  discrete tight-binding model. In that setting, it is straightforward to computationally analyze the dynamics of impinging excitonic packets because the Schr\"{o}dinger equation can be expressed as a set of coupled ordinary differential equations. In the simplest case, a single exciton packet can be made to oscillate between the two channels with an associated Rabi oscillation frequency. By adjusting the spatial footprint of the inter-channel coupling parameter, a 50/50 beam splitter can be realized. Two such beam splitters can then combined to produce a Mach-Zehnder interferometer.

In Section \ref{Ham}, we formulate the tight-binding model used to explore the excitonic dynamics. Section \ref{sec:osc} explains the mechanism of excitonic population transfer between the two conduit channels while Section \ref{sec:bs} uses the inter-channel oscillation to develop an excitonic beam splitter analogous to its optical counterpart. The beam splitter is then computationally simulated and shown to have the expected phase-shift properties\cite{Degiorgio_1980, Zeilinger_1981}. In Section \ref{sec:mzi}, we use the Hamiltonian formulation of the beam splitter to computationally simulate a Mach-Zehnder interferometer and reproduce the population output associated with excitonic wave-packet interference. 

\section{Tight-Binding Model}\label{Ham}
 
Consider a pair of identical one-dimensional lattices for which each site supports a single excited state. The associated Hamiltonian is then:
\begin{equation}
\hat H = \hat H_{{\rm site}} + \sum_{\nu = 1}^2 \hat H_\nu + \hat H_{{\rm cross}} . \label{Hall}
\end{equation}
The first term,
\begin{equation}
\hat H_{{\rm site}} = \sum_{j=1}^N \sum_{\nu = 1}^2 \Delta \hat c^{\nu \dagger}_j \hat c^\nu_j ,
\end{equation}
accounts for the energy, $\Delta$, of each site, $j$, on channel $\nu$. Here $\hat c^{\nu \dagger}_{j}$ is the creation operator for an exciton on site $j$ of channel $\nu$, and Bosonic commutation relations are obeyed:
\begin{eqnarray}
\bigl[ \hat c^\mu_i, \hat c^{\nu\dagger}_j \bigr] &=& \delta_{i,j} \delta_{\mu,\nu} \nonumber \\
\bigl[ \hat c^\mu_i, \hat c^{\nu}_j \bigr] &=& \bigl[ \hat c^{\mu\dagger}_i, \hat c^{\nu\dagger}_j \bigr]  = 0 . \label{comm_rels}
\end{eqnarray}
Operators $\hat H_1$ and $\hat H_2$ characterize the intra-chain hopping (delocalization) energies of each channel:
\begin{equation}
\hat H_\nu = \sum_{\braket{i, j}} \tau \hat c^{\nu \dagger}_{i} \hat c^\nu_{j} , \quad \nu = 1, 2.
\end{equation}
The last component of the Hamiltonian in Equation \ref{Hall} allows for symmetric exciton hopping between the two channels:
\begin{equation}
\hat H_{{\rm cross}} = \sum_{n} \chi_n \hat c^{1\dagger}_n \hat c^2_n + H.c.
\end{equation}
As with intrachain hopping, this is local in the sense that sites must be facing each other to interact, but here the hopping parameters, $\chi_n$, are now functions of position. 

\begin{figure}[hptb]
\begin{center}
\includegraphics[width=0.48\textwidth]{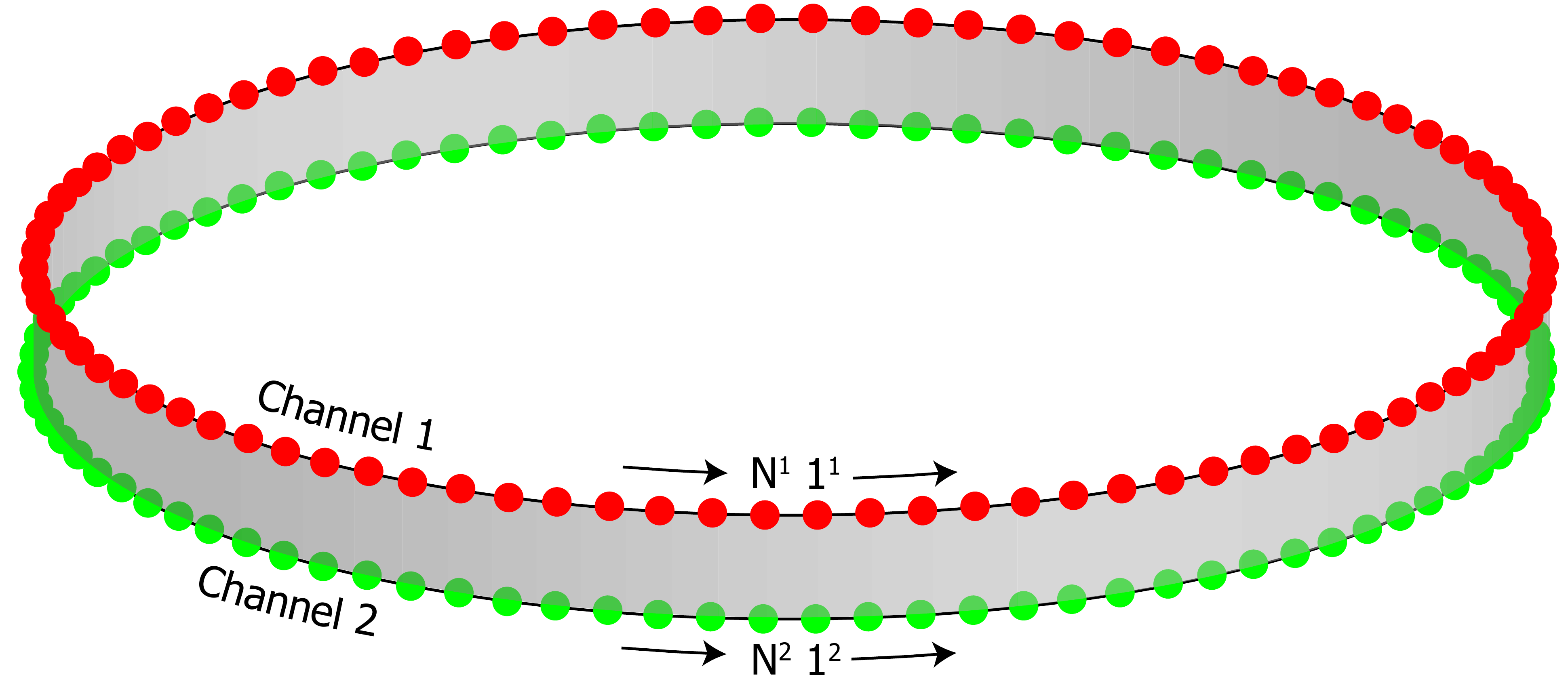}
\end{center}
\caption{\emph{Inductive excitonic beam splitter.} Periodic boundary conditions are enforced for both channels.}
\label{PBC}
\end{figure}

For the sake of analytical convenience, assume that both channels have periodic boundary conditions, as shown in Figure \ref{PBC}.
%
%
In the absence of any inter-channel coupling, each channel then obeys a simple dispersion relation,
\begin{equation}
\hbar \omega_j = \Delta + 2 \tau {\rm cos}(k_j a), \label{dispersion}
\end{equation}
where $k_j \equiv \frac{2\pi j}{N a}$ are the $N$ possible wavenumbers, $\omega_j$ are the associated temporal frequencies, and $a$ is the site spacing. This allows wave packets to be constructed from periodic eigenstates, endowed with a well-defined group velocity, and analyzed in the absence of boundary effects. The packet speed (group velocity) is given by
\begin{equation}
v_g(k) = \partial_k \omega(k) = \frac{-2 a \tau}{\hbar} {\rm sin}(k a) . \label{vgroup}
\end{equation}

Assume that the state of the system is an evolving, normalized superposition of excited states that collectively represent the result of a single excitation:
\begin{equation}
\ket{\Psi(t)}  = \sum_{j=1}^N \sum_{\nu = 1}^2 u^{(\nu)}_j(t) \hat{c}^{\nu \dagger}_{j}\ket{\rm vac} .
\label{initF}
\end{equation}
Here vacuum state, $\ket{{\rm vac}}$, is identified as the condition in which all sites are in their ground state.
The Schr\"{o}dinger equation then implies that the evolving state can be tracked by solving a set of 2N coupled ordinary differential equations for the quantum amplitudes of each ring site, $u^{(\nu)}_j$. The Channel 1 equations are

\begin{equation}
\imath \hbar \dot u^{(1)}_j =  \Delta u^{(1)}_j + \tau u^{(1)}_{j-1} + \tau u^{(1)}_{j+1} + \chi_j u^{(2)}_{j}
\label{ringODE}
\end{equation} 
with an analogous, coupled set for Channel 2. 

\section{Oscillation Between Channels}\label{sec:osc}

\begin{figure}[hptb]
\begin{center}
\includegraphics[width=0.48\textwidth]{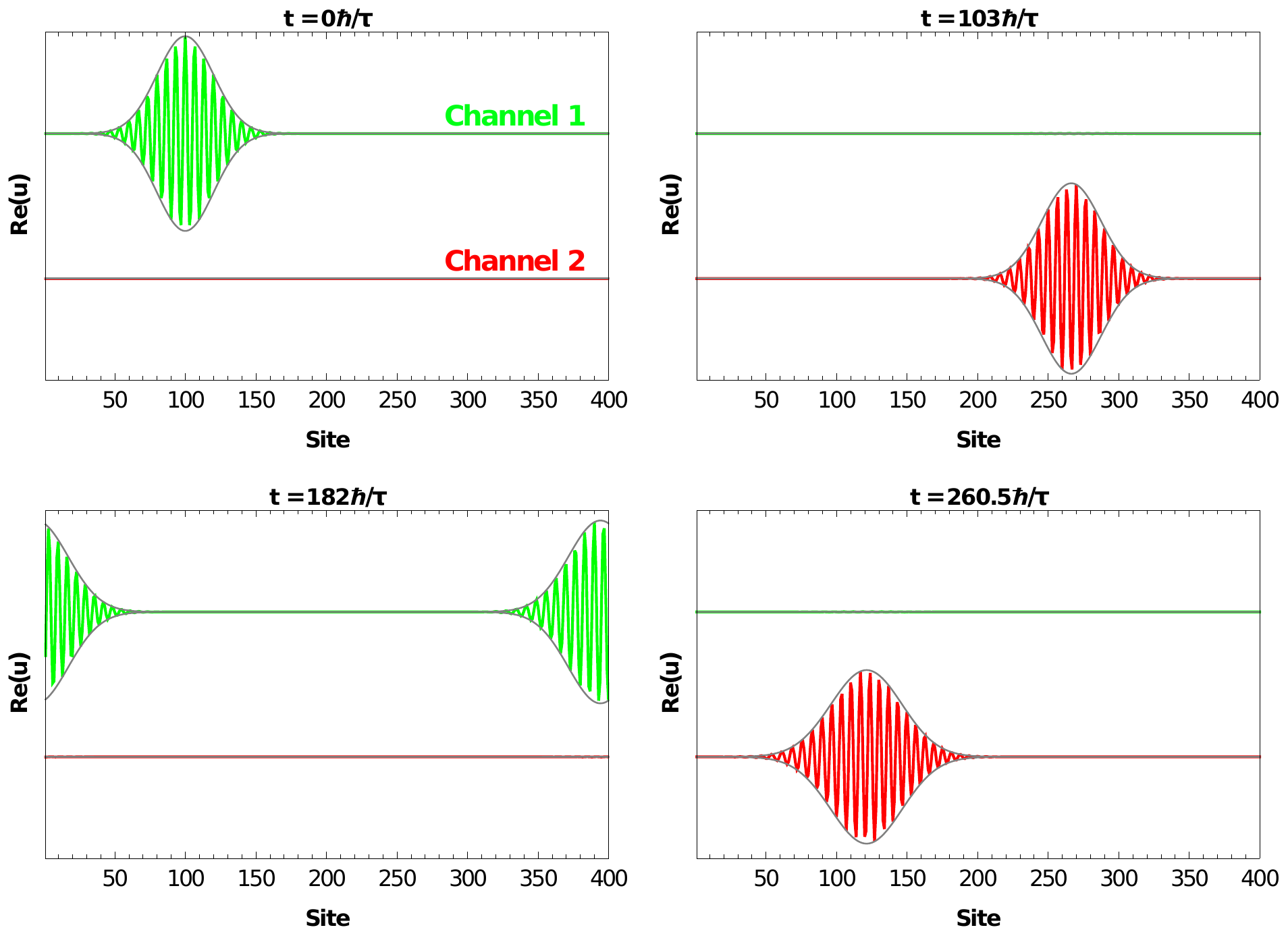}
\end{center}
\caption{\emph{Oscillation between channels.} The interaction, $\chi$, between channels is turned on smoothly at $t = 0$ as given by Equation \ref{chi_oscillation}. The associated parameters are: $\chi_0 = \tau/50$, $t_0 = 25 \hbar/\tau$, $k_0 a = 0.942$, and $\Delta = 0$. A Gaussian packet that is initially on Channel 1 subsequently oscillates between the channels at a rate determined by $\chi$. The green and red curves show the real part of the quantum amplitudes on each channel, while the solid gray envelopes are the magnitudes of the total amplitudes. A complete transfer of population from one channel to the next occurs over time interval $\pi/(2\chi)$.}
\label{Result_Oscillation}
\end{figure}

A wave packet that is initially traveling along Channel 1 will undergo a collective population oscillation between the channels as it propagates. This can be illustrated by turning on a uniform inter-channel interaction with an exponential switch:
\begin{equation}
\chi_j = \chi_0 e^{-t/t_0} .
\label{chi_oscillation}
\end{equation} 
The initial condition is taken to be a Gaussian wave packet moving to the right that is initially on Channel 1:
\begin{eqnarray}
u^{(1)}_j(0) &=& \frac{1}{\pi^{1/4}\sigma^{1/2}} e^{-\frac{(j - j_0)^2}{2\sigma^2}} e^{-\imath k_0 j} \nonumber \\
u^{(2)}_j(0) &=& 0. \label{IC_oscillation}
\end{eqnarray}
Here $\sigma$ is the standard deviation of the packet footprint, $j_0$ gives its position, and $k_0$ is a specified central wavenumber that controls the packet speed, Equation \ref{dispersion}.  The results are shown in Figure \ref{Result_Oscillation}, where it is clear that the exciton packet is transferred back and forth between channels as it moves. 

%
%
%
This rate of population transfer can be well-approximated by considering inter-channel dynamics in the absence of population flux from the left and right neighbors. A pair of ordinary differential equations then idealizes the transfer dynamics:
\begin{equation}
\imath \partial_t u_1 = \chi u_2, \quad \quad
\imath \partial_t u_2 = \chi u_1 ,
\label{chi_oscillation_diff}
\end{equation}
where $u_1$ and $u_2$ are the time-varying amplitudes on each channel. If the packet starts on Channel 1, then $u_1(t) = {\rm Cos}[\chi t]$ and $u_2(t) = -{\rm Sin}[\chi t]$. The populations thus cycle with a period of $T=\pi/\chi$, in excellent agreement with the results of Figure \ref{Result_Oscillation} once the oscillatory motion is established. This will be a good approximation provided that the net intra-channel population flux is much smaller than the flux between channels. The accuracy of this approximation therefore increases with the width of the packet and is inversely proportional to its speed.

In contrast to its optical counterpart, the excitonic beam splitter can involve more than two channels. If there are $N$ daughter channels, each with identical coupling to the parent channel and without coupling to each other, an extension of the analysis above shows that all $N+1$ channels will have equal populations at 
\begin{equation}
t_{\rm{eq}} = \frac{\rm{sec}^{-1}(\sqrt{N+1})}{\chi\sqrt{N}} .
\end{equation}
%
For instance, 3-channel beam-splitting could be carried out by having two daughter channels cross the parent channel at oblique angles of opposite sign from above and below.

\section{Beam-Splitting}\label{sec:bs}
The uniform, temporal, exponential switch is next replaced by an inter-channel coupling that is static in time and varies with position with a Gaussian distribution:
\begin{equation}
\chi_j = \chi_0 e^{-\frac{a(j - N/2)}{2 \sigma_\chi}} .
\label{chi_splitter}
\end{equation} 
Here $\sigma_\chi$ controls the spatial extent of the region in which the coupling exists. At issue is how to structure this inter-channel coupling so that 50\% of the population is transferred. The previous analysis of simple oscillation between two sites indicates that the populations are first equal to each other at $t = T/4$. At that moment, $u_1(T/4) = 1/\sqrt{2}$ and $u_2(T/4) = -\imath/\sqrt{2} = e^{-\imath \pi/2}u_1(T/4)$. This implies that a region in which the coupling, $\chi$, is constant will need to be of width $L = \frac{T |v_g(k_0)|}{4} = \frac{\pi |v_g(k_0)|}{4\chi}$. Such a sharply demarcated domain will result in the generation of reflections at either end though. This can be circumvented by noting that the system linearity only requires that the spatial integration of the function, $\rho(x)$, representing the coupling density be equal to $\frac{\pi |v_g(k_0)|}{4}$. Assuming a Gaussian coupling density with standard deviation $\sigma_\chi$,
\begin{equation}
\rho(x) = \chi e^{\frac{-x^2}{2\sigma^2_\chi}} ,
\label{chi_density}
\end{equation} 
a 50/50 beam division will occur provided $\int_{-\infty}^{\infty} d x\, \rho(x) = L \chi$. This implies that the Gaussian footprint of the coupling must have the following standard deviation:
\begin{equation}
\sigma_\chi = \frac{L}{\sqrt{2\pi}} = \frac{\pi |v_g(k_0)|}{4\chi\sqrt{2\pi}}  = \frac{a\tau \sqrt{2\pi}}{4\chi\hbar} |{\rm sin}(k a)|.
\label{sigma_chi}
\end{equation} 
The simple analysis of population oscillation also indicates that such a beam splitting process will result in a phase shift of $-\pi/2$ on Channel 2. This strategy for \emph{inductively splitting} an excitonic wave packet is essentially the same as that originally employed to transfer excited state energy from ammonia molecules in an microwave cavity for masers~\cite{Townes_1955}.

\begin{figure}[hptb]
\begin{center}
\includegraphics[width=0.48\textwidth]{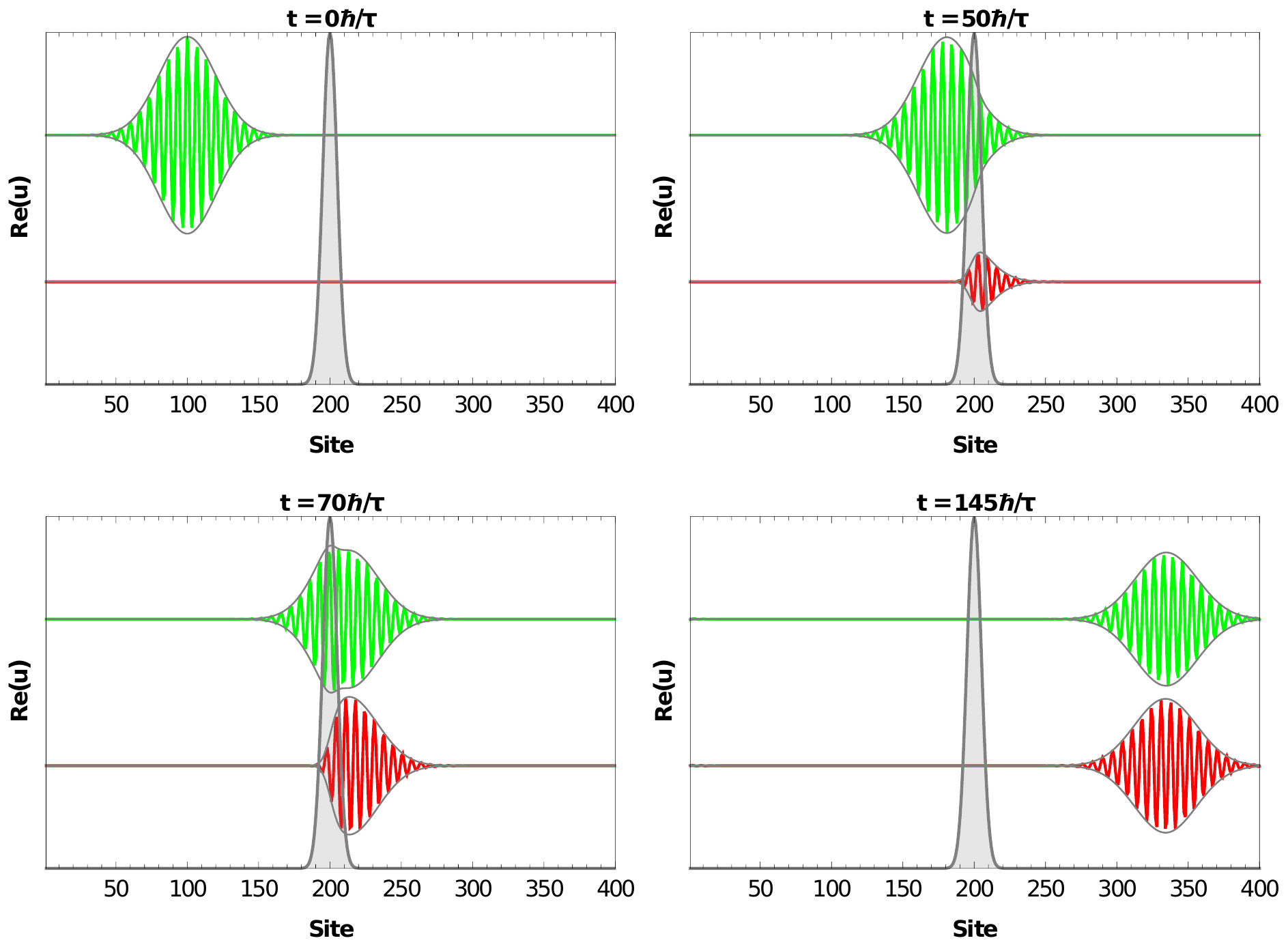}
\end{center}
\caption{\emph{Beam splitting between channels.} A Gaussian packet on Channel 1 moves into a region in which there is coupling between the channels. As a result, the packet splits into two identical packets that are phase shifted by $\pi/2$. The green and red curves show the real part of the quantum amplitudes on each channel, while the solid gray enveloping curves give the associated magnitude of the difference. The coupling between channels, shown as a filled gray shape, is static and has a spatial distribution centered at $N/2$. The associated parameters are: $\sigma = 20 a$, $\chi_0 = \tau/10$, $\sigma_\chi = 5.07 a$, $k_0 a = 5.34$, and $\Delta = 0$.} 
\label{Result_Beam_Splitter}
\end{figure}

\begin{figure}[hptb]
\begin{center}
\includegraphics[width=0.48\textwidth]{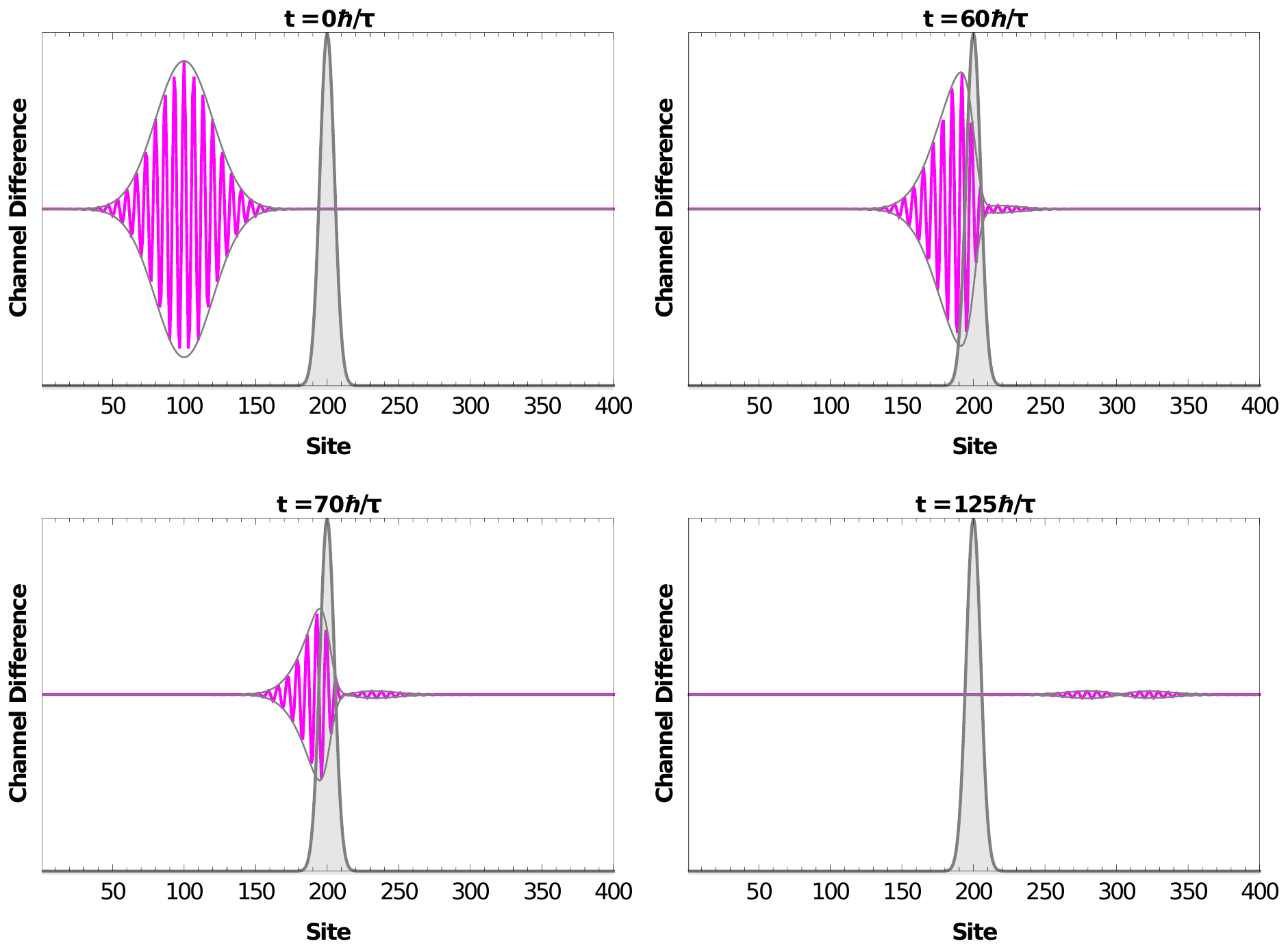}
\caption{\emph{Phase shift in the two-channel beam splitter.} The amplitudes of Figure \ref{Result_Beam_Splitter} are plotted as the difference between the two channels, with the Channel 2 data pre-multiplied by a factor of $e^{\imath \pi/2}$.  The magenta curve shows the real part of this amplitude difference, while the solid gray enveloping curves give the associated magnitude of the difference. The coupling between channels, shown as a filled gray shape, is static and has a spatial distribution centered at $N/2$.} 
\label{Result_Diff}
\end{center}
\end{figure}

The approach is computationally shown in Figure \ref{Result_Beam_Splitter}, where a Gaussian wave packet on Channel 1 is divided into two identical packets. A quantitative comparison of the resulting packets is obtained by phase shifting the second one by $\pi/2$ and subtracting it from the first. This is shown in Figure \ref{Result_Diff}. 

%
%

As in the 50/50 dielectric optical beam splitter~\cite{Degiorgio_1980, Zeilinger_1981}, there is a $\pi/2$ phase shift between the channel packets. Interestingly, it is the Schr\"{o}dinger equation, and not the Fresnel coefficients derived from Maxwell's equations, that are the source of this shift for excitons.

%
%

\section{Mach-Zehnder Interferometer}\label{sec:mzi}

\begin{figure}[hptb]
\begin{center}
\includegraphics[width=0.48\textwidth]{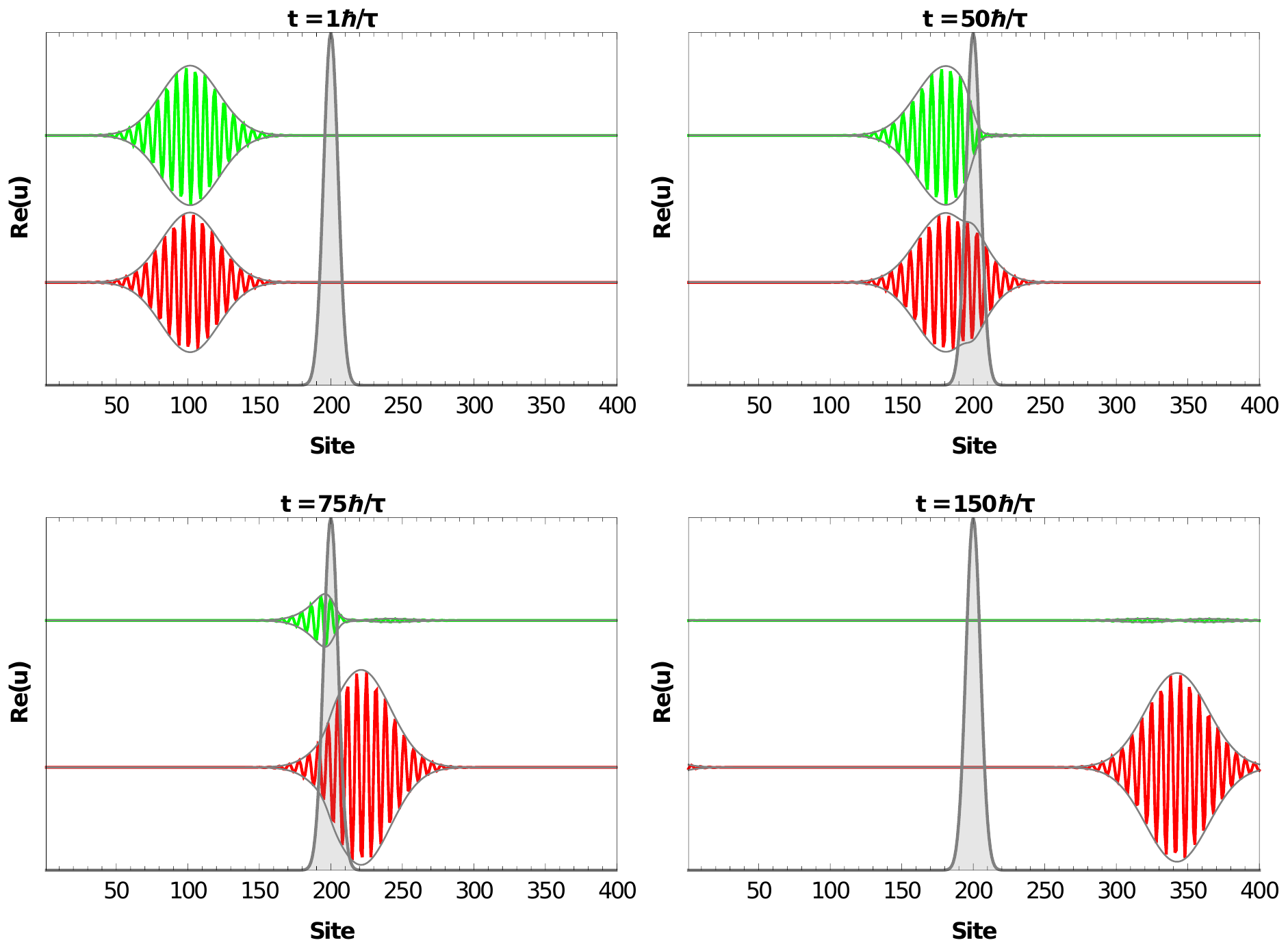}
\caption{\emph{Constructive/destructive interference resulting in beam annihilation.} Identical Gaussian packets, that differ only in that the signal of Channel 2 is phase shifted by $-\pi/2$ phase shift, move into a region in which there is coupling between the channels. As a result, the packets interfere constructively on Channel 1 and destructively on Channel 2. This amounts to an annihilation of the packet on Channel 2. The green and red curves show the real part of the quantum amplitudes on each channel, while the solid gray enveloping curves give the associated magnitude of the difference. The coupling between channels, shown as a filled gray shape, is static and has a spatial distribution centered at $N/2$.  The associated parameters are: $\sigma = 20 a$, $\chi_0 = \tau/10$, $\sigma_\chi = 5.07 a$, $k_0 a = 5.34$, and $\Delta = 0$.} 
\label{Result_Interferometry}
\end{center}
\end{figure}

Optical experiments often involve interferometric measurements between two beams of light by using a beam splitter in which the input channels are allowed to interfere. The simplest case of this amounts to a time reversal of the beam splitter operation of Figure \ref{Result_Beam_Splitter} in which two beams that are identical, except that the Channel 2 packet is phase shifted by $-\pi/2$, combine destructively on one line and constructively on the other. This is demonstrated in Figure \ref{Result_Interferometry} and simulates the effect of sending a Channel 1 packet through a sequence of two beam splitters resulting in a complete transfer of population to Channel 2. The result suggests that the introduction of a phase shift on one leg could be used as a basis for carrying out excitonic interferometry experiments. 

\begin{figure}[hptb]
\begin{center}
\includegraphics[width=0.48\textwidth]{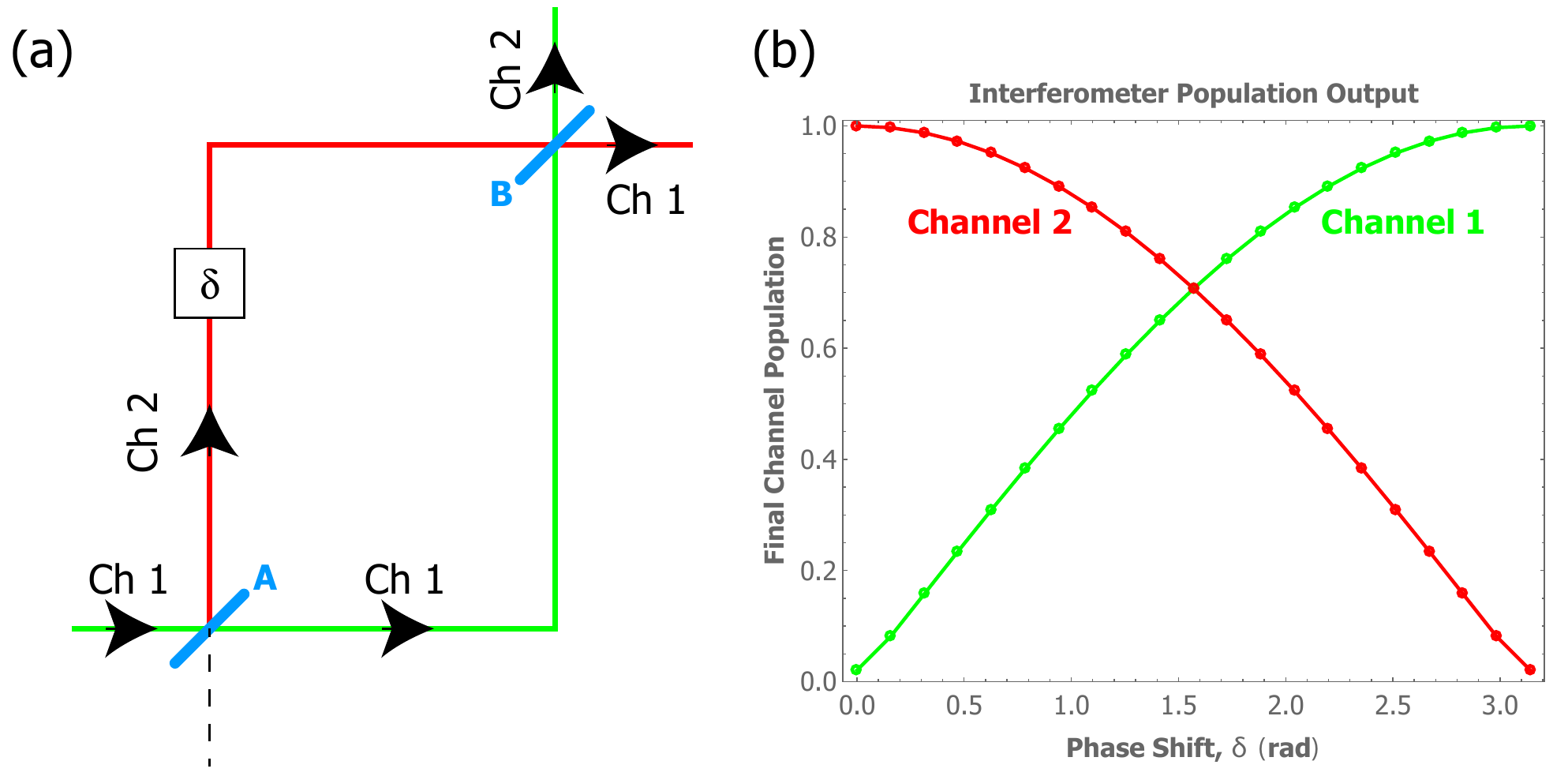}
\end{center}
\caption{\emph{Excitonic Mach-Zehnder Interferometer.} Panel (a): The beam splitters of Figures \ref{Result_Beam_Splitter} and \ref{Result_Interferometry} are combined along with a tunable phase shift, $e^{\imath \delta}$, on Channel 2. Interference between the two channels can be controlled by adjusting the phase shift, $\delta$. Beam splitters are shown in blue. Panel (b): The final population of the channels is plotted as a function of the phase shift, $\delta$, demonstrating that the population on each channel can be controlled.  The associated parameters are: $\sigma = 20 a$, $\chi_0 = \tau/10$, $\sigma_\chi = 5.07 a$, $k_0 a = 5.34$, and $\Delta = 0$.
} 
\label{Result_Phaseshift}
\end{figure}

Indeed, an excitonic version of the Mach-Zehnder interferometer~\cite{Zehnder_1891, Mach_1892} can be constructed by sequentially combining the beam splitters of Figures \ref{Result_Beam_Splitter} and \ref{Result_Interferometry} along with a tunable phase shift, $e^{\imath \delta}$, on Channel 2. The associated schematic is shown in the panel (a) of Figure \ref{Result_Phaseshift}, while panel (b) shows how the final state populations can be arbitrarily tuned using the phase shift. This further implies such an excitonic interferometer can be used to quantify the temporal and/or spatial coherence between two packets.

%
%

%
%

\section{Discussion}

Exciton wave packets can be induced to travel down conduits due to intra-channel coupling between neighboring sites, but they can also be transferred between such conduits where there also exists a spatially-dependent inter-channel coupling. In actual implementations, this can be as simple as varying the separation between the conduits. Within a tight-binding setting, it has been demonstrated that fabricating inter-channel coupling with a Gaussian spatial distribution amounts to an inductive excitonic beam splitter analogous to its optical counterpart. Unlike in the case of a simple optical beam splitter, it is possible to design multi-channel excitonic beam splitters. 

An immediate extension of this design philosophy results in a Mach-Zehnder interferometer, enabling excitonic interferometry experiments. As just one example, the combination of the interferometer with a previously designed excitonic version of spontaneous parametric downconversion\cite{excbell} makes it possible to carry out an excitonic version of delayed-choice quantum erasure.

A number of simplifying assumptions were made that idealize the physical setting in which excitonic components are most likely to be realized. In order to focus on beam-splitting character, the model assumes that  excitonic wavepackets are not coupled to their environment--e.g. to phonons in the system~\cite{PhysRevB.95.195423}.  If such coupling were to be taken into account, it would be possible to explore entanglement between excitons in the presence of dephasing~\cite{PhysRevB.92.241112}.

Another simplifying assumption is that the tight-binding model takes each exciton to be localized at a single site--i.e. an assumption of Frenkel excitons as opposed to spatially diffuse Wannier-Mott excitons. It is further assumed that the exciton binding energy is sufficiently high that the excitons do not disassociate into constituent electrons and holes. Finally, only nearest-neighbor interactions are considered.

The exciton beam splitter can be experimentally realized in a number of manifestations. For instance, the channels could be molecular chains with each link exhibiting weak coupling to its neighbors and an energy structure with the first excited state widely separated from higher energy levels. A second possible manifestation would be to realize each site as an optical cavity likewise coupled to its neighbors.

\bibliographystyle{prsty}
\bibliography{Beam_Splitter}{}

\begin{thebibliography}{10}

\bibitem{PhysRevLett.89.097401}
H. Zhao, S. Moehl, and H. Kalt, Phys. Rev. Lett. {\bf 89},  097401  (2002).

\bibitem{PhysRevB.95.195423}
X. Zang, S. Montangero, L.~D. Carr, and M.~T. Lusk, Phys. Rev. B {\bf 95},
  195423  (2017).

\bibitem{Dostál2018}
J. Dost{\'a}l, F. Fennel, F. Koch, S. Herbst, F. W{\"u}rthner, and T. Brixner,
  Nature Communications {\bf 9},  2466  (2018).

\bibitem{Degiorgio_1980}
V. Degiorgio, American Journal of Physics {\bf 48},  81  (1980).

\bibitem{Kim2013}
H. Kim, R. Bose, T.~C. Shen, G.~S. Solomon, and E. Waks, Nature Photonics {\bf
  7},  373 EP   (2013).

\bibitem{PhysRevA.93.062342}
S. Seifnashri, F. Kianvash, J. Nobakht, and V. Karimipour, Phys. Rev. A {\bf
  93},  062342  (2016).

\bibitem{PhysRevA.69.052315}
T.~J. Osborne and N. Linden, Phys. Rev. A {\bf 69},  052315  (2004).

\bibitem{PhysRevA.72.062326}
H.~L. Haselgrove, Phys. Rev. A {\bf 72},  062326  (2005).

\bibitem{doi:10.1080/00107510701342313}
S. Bose, Contemporary Physics {\bf 48},  13  (2007).

\bibitem{PhysRevLett.91.207901}
S. Bose, Phys. Rev. Lett. {\bf 91},  207901  (2003).

\bibitem{PhysRevA.65.052313}
N.-K. Tran and O. Pfister, Phys. Rev. A {\bf 65},  052313  (2002).

\bibitem{Bouwmeester1997}
D. Bouwmeester, J.-W. Pan, K. Mattle, M. Eibl, H. Weinfurter, and A. Zeilinger,
  Nature {\bf 390},  575  (1997).

\bibitem{PhysRevLett.76.4656}
K. Mattle, H. Weinfurter, P.~G. Kwiat, and A. Zeilinger, Phys. Rev. Lett. {\bf
  76},  4656  (1996).

\bibitem{Dong2008}
R. Dong, M. Lassen, J. Heersink, C. Marquardt, R. Filip, G. Leuchs, and U.~L.
  Andersen, Nature Physics {\bf 4},  919 EP   (2008).

\bibitem{PhysRevLett.84.1}
Y.-H. Kim, R. Yu, S.~P. Kulik, Y. Shih, and M.~O. Scully, Phys. Rev. Lett. {\bf
  84},  1  (2000).

\bibitem{PhysRevLett.92.210403}
J.~C. Howell, R.~S. Bennink, S.~J. Bentley, and R.~W. Boyd, Phys. Rev. Lett.
  {\bf 92},  210403  (2004).

\bibitem{PhysRevA.85.032121}
R. Auccaise, R.~M. Serra, J.~G. Filgueiras, R.~S. Sarthour, I.~S. Oliveira, and
  L.~C. C\'eleri, Phys. Rev. A {\bf 85},  032121  (2012).

\bibitem{excbell}
M.~T. Lusk, {arXiv}:1706.07420, (2017).

\bibitem{PhysRevB.96.155104}
X. Zang and M.~T. Lusk, Phys. Rev. B {\bf 96},  155104  (2017).

\bibitem{Kasprzak2006}
J. Kasprzak, M. Richard, S. Kundermann, A. Baas, P. Jeambrun, J.~M.~J. Keeling,
  F.~M. Marchetti, M.~H. Szymanska, R. Andr{\'e}, J.~L. Staehli, V. Savona,
  P.~B. Littlewood, B. Deveaud, and L.~S. Dang, Nature {\bf 443},  409  (2006).

\bibitem{doi:10.1021/nl300983n}
A.~A. High, J.~R. Leonard, M. Remeika, L.~V. Butov, M. Hanson, and A.~C.
  Gossard, Nano Letters {\bf 12},  2605  (2012), pMID: 22509898.

\bibitem{doi:10.1088/1361-6455}
R.~S. Heinrich~Stolz, Maria~Dietl and D. Semkat, Journal of Physics B: Atomic,
  Molecular and Optical Physics {\bf 51},    (2017).

\bibitem{PhysRevB.92.241112}
M.~T. Lusk, C.~A. Stafford, J.~D. Zimmerman, and L.~D. Carr, Phys. Rev. B {\bf
  92},  241112  (2015).

\bibitem{1253058}
X. {Lei}, S. {Bai}, H. {Ge}, and S.~Y. {Chou},  in {\em The 16th Annual Meeting
  of the IEEE Lasers and Electro-Optics Society, 2003. LEOS 2003.} (IEEE,
  ADDRESS, 2003), Vol.~2, pp.\ 832--833 vol.2.

\bibitem{J_ns_2015}
K.~D. Jöns, U. Rengstl, M. Oster, F. Hargart, M. Heldmaier, S. Bounouar, S.~M.
  Ulrich, M. Jetter, and P. Michler, Journal of Physics D: Applied Physics {\bf
  48},  085101  (2015).

\bibitem{Zeilinger_1981}
A. Zeilinger, American Journal of Physics {\bf 49},  882  (1981).

\bibitem{Townes_1955}
J.~P. Gordon, H.~J. Zeiger, and C.~H. Townes, Phys. Rev. {\bf 99},  1264
  (1955).

\bibitem{Zehnder_1891}
L. Zehnder,  in {\em Zeitschrift für Instrumentenkunde}, edited by E. Dorn
  (Physikalisch-Technische Reichsanstalt, Germany, 1891), pp.\ 275--285.

\bibitem{Mach_1892}
L. Mach,  in {\em Zeitschrift für Instrumentenkunde}, edited by E. Dorn
  (Physikalisch-Technische Reichsanstalt, Germany, 1892), pp.\ 89--93.

\end{thebibliography}

\end{document}